\documentclass[twocolumn,twocolappendix, trackchanges]{aastex631}

\usepackage{amsmath}
\usepackage{multirow}

\shorttitle{Initial spins of neutron stars}
\shortauthors{Du et al.}

\graphicspath{{./}{Figures/}}

\begin{document}

\title{On the initial spin period distribution of neutron stars}

\author[0000-0002-0986-218X]{Shen-Shi Du}
\affiliation{School of Physics and Technology, Wuhan University, Wuhan, Hubei 430072, China}
\affiliation{Department of Physics, Faculty of Arts and Sciences, Beijing Normal University, Zhuhai 519087, China}
\affiliation{Advanced Institute of Natural Sciences, Beijing Normal University, Zhuhai 519087, China}

\author[0000-0002-2187-4087]{Xiao-Jin Liu}
\affiliation{Department of Physics, Faculty of Arts and Sciences, Beijing Normal University, Zhuhai 519087, China}
\affiliation{Advanced Institute of Natural Sciences, Beijing Normal University, Zhuhai 519087, China}
\affiliation{Department of Astronomy, Beijing Normal University, Beijing 100875, China}

\author[0000-0001-7016-9934]{Zu-Cheng Chen}
\affiliation{Advanced Institute of Natural Sciences, Beijing Normal University, Zhuhai 519087, China}
\affiliation{Department of Astronomy, Beijing Normal University, Beijing 100875, China}
\affiliation{Department of Physics and Synergetic Innovation Center for Quantum Effects and Applications, Hunan Normal University, Changsha, Hunan 410081, China}
\affiliation{Institute of Interdisciplinary Studies, Hunan Normal University, Changsha, Hunan 410081, China}

\author[0000-0002-3309-415X]{Zhi-Qiang You}
\affiliation{Advanced Institute of Natural Sciences, Beijing Normal University, Zhuhai 519087, China}
\affiliation{Department of Astronomy, Beijing Normal University, Beijing 100875, China}
\affiliation{Henan Academy of Sciences, Zhengzhou 450046, Henan, China}

\author[0000-0001-7049-6468]{Xing-Jiang Zhu}
\affiliation{Department of Physics, Faculty of Arts and Sciences, Beijing Normal University, Zhuhai 519087, China}
\affiliation{Advanced Institute of Natural Sciences, Beijing Normal University, Zhuhai 519087, China}
\affiliation{Institute for Frontier in Astronomy and Astrophysics, Beijing Normal University, Beijing 102206, China}

\author{Zong-Hong Zhu}
\affiliation{School of Physics and Technology, Wuhan University, Wuhan, Hubei 430072, China}

\correspondingauthor{Xing-Jiang Zhu}
\email{zhuxj@bnu.edu.cn}
\correspondingauthor{Zong-Hong Zhu}
\email{zzh@whu.edu.cn}

\begin{abstract}
We derive the initial spin period distribution of neutron stars by studying the population of young pulsars associated with supernova remnants. Our hierarchical Bayesian approach accounts for the measurement uncertainties of individual observations and selection effects. 
Without correcting for selection effects, as done in previous studies, we find that pulsar initial spin periods follow a Weibull distribution, peaking at 40~ms, which is favoured against the lognormal distribution with a Bayes factor of 200. 
The known selection effects in radio pulsar surveys, 
including pulse broadening and a period-dependent beaming fraction, have been quantitatively investigated.
We show that, based on measurements of pulsar luminosity and spin period 
from the ATNF Pulsar Catalogue, 
the impact of pulse broadening on the inference of pulsar period distribution 
is likely to be insignificant. 
Correcting for the beaming selection effect, a Weibull distribution remains to be the preferred model, while its peak slightly shifts to longer periods at 50~ms. 
Our method will prove useful in constraining the birth properties of neutron stars in the Square Kilometre Array era. 

\end{abstract}


\section{Introduction} \label{sec:sec1}
Pulsars are fast rotating, highly magnetized neutron stars (NSs). 
Since the first discovery \citep{Hewish1968Natur.217..709H}, 
the number of observed pulsars has grown to over 3500 \citep[see the ATNF Pulsar Catalogue\footnote{\url{https://www.atnf.csiro.au/research/pulsar/psrcat/}},][]{MHT+05}, a majority of which are detected in the radio band.
Among them, pulsars observed in association with supernova remnants (SNRs) are of particular interest, since their ages are independently informed by observations of SNRs.
The Crab pulsar is the best known example.
Firmly established as the remnant star of supernova 1054, its initial spin period is around 20~ms, close to its current spin period of 33~ms.
This is widely used as a proxy for pulsar initial spin periods in the community \citep[e.g.,][]{Johnston17PPdot}.

The astrophysical processes that give rise to NS spins are poorly understood.
A range of spin periods from milliseconds to seconds are predicted in a variety of processes during supernova explosions. 
Newborn NSs could inherit the angular momentum of progenitor stars from the collapsing iron cores \citep{Heger2005ApJ...626..350H, Ott2006ApJS..164..130O}, where the angular momentum transport and mass loss in single stars play a significant role \citep{Fuller2014ApJ...796...17F, Fuller2015MNRAS.450..414F, Fuller2019MNRAS.485.3661F, Ma2019MNRAS.488.4338M, 
Eggenberger2019A&A...631L...6E, Hu2023arXiv230106402H}. 
For instance, \cite{Ott2006ApJS..164..130O} found that an NS could be born with periods of tens to hundreds of milliseconds if the spin periods of iron cores are around 50$-$100~s. 
NS spins could also stem from the natal kicks caused by the asymmetric mass ejection and anisotropic neutrino emission (e.g.,  \citealt{Spruit1998Natur.393..139S, Ng2007ApJ...660.1357N, Janka2022ApJ...926....9J, Coleman2022MNRAS.517.3938C, Fragione2023arXiv230508920F, Burrows2023arXiv231112109B}). 
In particular, \cite{Spruit1998Natur.393..139S} showed that an off-center kick at a speed of several hundreds~km$/$s could lead to a spin period as short as tens of milliseconds. 
Other processes that might play a role in producing NS spins include hydrodynamic instabilities during the supernova explosions such as the standing accretion shock instability (e.g., \citealt{Blondin2007Natur.445...58B, Guilet2014MNRAS.441.2782G, Kazeroni2016MNRAS.456..126K}),
the anisotropic accretion of angular momentum during the pre-explosion phase \citep{Wongwathanarat2013A&A...552A.126W} and post-explosion phase \citep[e.g.,][]{Janka2022ApJ...926....9J}. 
Therefore, the initial period distribution of NSs is a powerful probe into astrophysical processes during their formation.

Thus far, around a hundred pulsars have been detected in association with SNRs. 
They are considered to be young (with a typical age less than $\sim10^5$~years), 
and therefore provide useful insights into the birth properties of NSs, such as the spin period, magnetic field, spatial velocity, and inclination angle between the spin and magnetic axes \citep{Popov2012Ap&SS.341..457P, Malov2021MNRAS.502..809M, Igoshev2022MNRAS.tmp.1595I}. 
To properly derive the initial distribution of pulsar parameters with the observed pulsar-SNR population, two approaches can be used.
First, by performing population syntheses \citep{Emmering1989ApJ...345..931E, FK2006ApJ...643..332F, Bates2014MNRAS.439.2893B, Gullon2014MNRAS.443.1891G, Cieslar2020MNRAS.492.4043C, Dirson2022A&A...667A..82D}, one develops a model pulsar population based on plausible assumptions about their initial properties and time evolution, and then runs the synthetic population through some mock pulsar surveys with an aim to reproduce the observed pulsar sample.
The model that best reproduces observations is considered to be a good model.
In the second approach, one extrapolates, for example, the initial periods of pulsars back from their measured spins and age estimates with assumptions about spin evolution (e.g., \citealt{Xu2023ApJ...947...76X}). 
Then, statistical inference is performed to fit the distribution of initial periods by incorporating selection effects and measurement uncertainties in individual observations. 
The major challenge in this approach is an appropriate quantification of the possible selection effects that would bias the observed distribution from the true population (e.g., \citealt{Lorimer2006MNRAS.372..777L, Lorimer2015MNRAS.450.2185L}). 
\cite{Igoshev2022MNRAS.tmp.1595I} performed maximum-likelihood estimates for the distributions of initial periods and magnetic fields of NSs using a sample of Galactic pulsars detected in association with SNRs. 
However, the selection effects were only qualitatively discussed. 

In this work, we revisit the inference of the initial period distribution of NSs.
For Galatic pulsars detected within SNRs, we 
derive the initial period of each pulsar from the measured spins, using the SNR age and assuming a generic spin-down model.
We then fit the distribution of initial spin periods through hierarchical Bayesian inference. 
Our analysis accounts for the uncertainties in age estimates of SNRs, effects of different braking indices, and selection effects in radio pulsar surveys.
We also perform Bayesian model selection to determine the best functional form of the initial period distribution. 

This paper is organized as follows.
In Section~\ref{sec:sec2}, we describe the observed sample, our data selection criteria and the spin-down evolution of radio pulsars. 
Section~\ref{sec:sec3.1} presents the statistical framework. 
The results and discussions of various effects that might affect our results are described in Section~\ref{sec:sec3.2}. 
We present concluding remarks in Section \ref{sec:sec4}.

\section{Data and Model}\label{sec:sec2}
In this section, we first describe the compilation of pulsars observed within SNRs and our data selection criteria. 
Subsequently, we present the model of spin-down evolution of pulsars. 
\subsection{Pulsars in supernova remnants}\label{sec:sec2.1}
Recently, \cite{Igoshev2022MNRAS.tmp.1595I} compiled $68$ pulsars observed in possible association with SNRs
by matching the Galactic SNR Catalogue\footnote{\url{ http://snrcat.physics.umanitoba.ca/SNRtable.php}} 
\citep{Ferrand2012AdSpR..49.1313F} with the ATNF Pulsar Catalogue. 
In this work, we add $20$ Galactic pulsar-SNR pairs to the sample adopted by \cite{Igoshev2022MNRAS.tmp.1595I}. 
These sources can also be found in the Galactic SNR Catalogue (SNRcat) and have been analyzed in previous works \citep{Popov2012Ap&SS.341..457P, Malov2021MNRAS.502..809M}.
In 
Appendix \ref{sec:append1}, we list these 88 Galactic pulsar-SNR pairs together with their parameters.
The association of the Monogem Ring with PSR J0659+1414, which is uncertain in the SNRcat, has recently been confirmed by \cite{Yao2022ApJ...939...75Y}. 

To derive the initial spin period ($P_0$) of pulsars from observed spin parameters and their age estimates by assuming a spin-down model, we select pulsars in our sample based on the following criteria. 
The pulsars must have their period ($P$) and period derivative ($\dot{P}$) measured. 
The values of $P$ and $\dot{P}$ are taken from the ATNF Catalogue (version 1.70). 
We assume that the age of SNR ($\tau_{\rm snr}$), taken from SNRcat, is the pulsar's true age unless otherwise specified (see Section \ref{sec:sec2.2} for further discussion). 
For PSR J1801$-$2451, the SNR age is unavailable, thus we adopt the kinematic age determined based on the measurements of its proper motion and position \citep{Noutsos2013MNRAS.430.2281N}.

The pulsar should be uniquely paired with the SNR. 
Since most of these pulsar-SNR sources are located near the Galactic plane,
there might be some overlaps between them by chance. 
Specifically, there are two issues: (1) an isolated pulsar is located close to multiple SNRs and (2) an isolated SNR is located close to multiple pulsars. 
We tackle the first issue by only selecting the pairs of which the SNR age is closest to the characteristic age of pulsar. 
Therefore, the pulsar J1640$-$4631 will be identified to be associated with SNR G338.3$-$00.0. 
To address the second issue, we select the pulsar whose characteristic age is closest to the SNR age. 
For example, in the case of SNR G035.6$-$00.4, three pulsars, J1857$+$0143, J1857$+$0210, and J1857$+$0212, were detected close to this remnant; 
we select PSR J1857$+$0143 as in association with G035.6$-$00.4. 
The included and excluded pairs are labeled as `Y' and `N' in the column ``Included" of the table in Appendix~\ref{sec:append1}, respectively.

Following \cite{Igoshev2022MNRAS.tmp.1595I}, we exclude magnetars in our sample. 
Thirteen Galactic pulsars in SNRs are known as magnetars (see 
Appendix~\ref{sec:append1}
), which have been reported in the McGill Online Magnetar Catalogue\footnote{\url{http://www.physics.mcgill.ca/~pulsar/magnetar/main.html}}. 
Magnetars could represent a distinct population different from normal radio pulsars (e.g., \citealt{Gullon2015MNRAS.454..615G}). The energy source powering magnetars could be more complicated than that of rotation-powered pulsars; the dominant energy loss could be in the forms of, e.g., the decay of magnetar's immense dipole field \citep{Duncan1992ApJ...392L...9D} and 
the significant gravitational radiation due to their large magnetic deformation and (initially) fast rotation \citep{Ioka2001MNRAS.327..639I, Dall'Osso2009MNRAS.398.1869D}.
Regardless of the loss mechanism, magnetars are thought to experience much more efficient spin deceleration before the onset of dipole spin-down, with substantially strong magnetic fields and a wide range of initial periods (e.g., \citealt{Thompson2004ApJ...611..380T, Prasanna2022MNRAS.517.3008P}).

\begin{figure}
\centering
\includegraphics[width=\columnwidth]{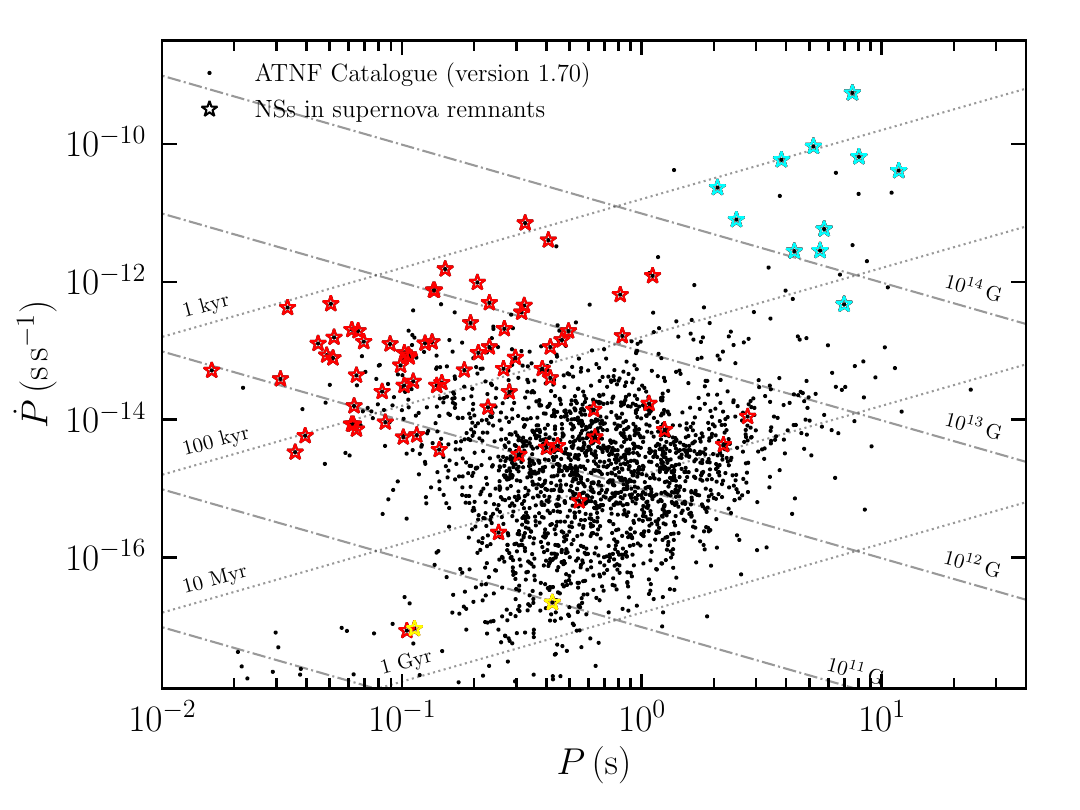}
\caption{
The $P$-$\dot{P}$ diagram of the observed pulsars. The stars denote the pulsars detected within SNRs, with CCOs and magnetars highlighted in yellow and cyan, respectively. The age and magnetic field strength ($B_{\rm dip}$) are calculated in the standard magneto-dipole model. 
The data set of this work is available at \url{https://github.com/Shen-Shi/NS_Birth_Population}.
}
 \label{MyFigA}
\end{figure}

Two pulsars, J0821$-$4300 and J1210$-$5226, known as central compact objects (CCOs), are included in our sample. 
The detection of X-ray pulsations provides strong support for the argument that they are pulsars \citep{Hui2006A&A...454..543H, Gotthelf2013ApJ...765...58G}. 
The period derivatives of these two objects are measured with phase-coherent X-ray timing by incorporating the measurements of position and proper motion \citep{Gotthelf2013ApJ...765...58G}. 
Note that \cite{Igoshev2022MNRAS.tmp.1595I} did not include the CCOs in their analysis of the initial period. 

We also require SNR age less than the characteristic age of pulsars, see the discussions in section~\ref{sec:sec2.2}. After applying the aforementioned selection criteria, we are left with $39$ pulsar-SNR pairs available for the inference of initial spin period distribution; see Appendix~\ref{sec:append1} for details.

\begin{figure}
\centering
\includegraphics[width=\columnwidth]{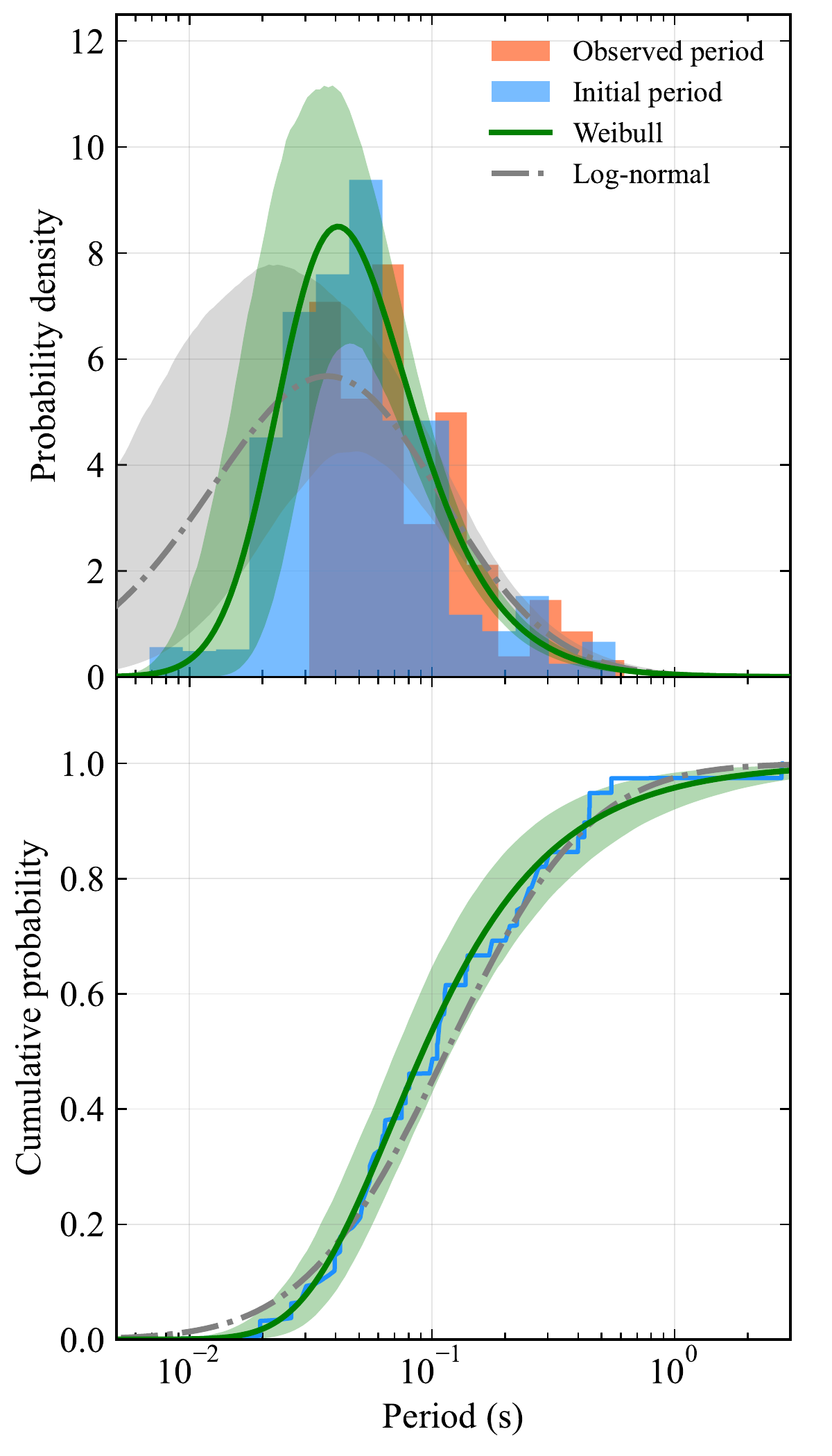}
\caption{
Upper panel: The distributions of the observed spin periods (orange) and initial periods (blue) of pulsars associated with SNRs. 
The green (gray) line denotes the posterior predictive distribution of initial period derived 
from the Weibull (log-normal) distribution (see section~\ref{sec:sec3.2.1}), with the shaded area indicating the 90\% credible interval. 
Lower panel: The cumulative distributions of the initial periods and the Weibull and log-normal model are plotted with the same colors as the upper panel.
}
\label{MyFig2}
\end{figure}

We show the pulsar $P$-$\dot{P}$ diagram in Figure~\ref{MyFigA}. The NSs associated with SNRs are denoted by stars: 
magnetars (cyan) populate the slowest and strongest magnetic filed regions among all species of neutron stars; 
normal pulsars and two CCOs are highlighted in red and yellow, respectively. 
In the upper panel of Figure~\ref{MyFig2}, we show the observed spin period distribution of the pulsars (orange histogram) included in our sample.

\subsection{Spin-down evolution of pulsars}\label{sec:sec2.2}
The pulsar spin-down evolution can be modeled with the following torque equation \citep{Manchester1977puls.book.....M}:
\begin{equation}\label{eq:sd}
  \dot{\Omega} \propto -\Omega^n,
\end{equation}
where $\Omega$ and $\dot{\Omega}$ are the spin angular frequency and
its first time derivative, respectively.
These two quantities can be accurately measured from pulsar timing observations \citep{Groth1975ApJS...29..453G}. 
The braking index is defined as $n = \Omega\ddot{\Omega}/\dot{\Omega}^2$,
where $\ddot{\Omega}$ is the second time derivative of spin angular frequency.
For a constant-field dipole in a vacuum, i.e., in the standard magneto-dipole model \citep{Ostriker1969ApJ...157.1395O}, the predicted value of $n$ is equal to $3$. 
A value (significantly) differing from $n=3$ indicates more complicated and separate braking processes in the rotational evolution, e.g., an evolving dipolar field (e.g., \citealt{Gao2017ApJ...849...19G}
and references therein), angular momentum loss through particle wind flows
\citep{Harding1999ApJ...525L.125H}, or gravitational radiation due to multipoles (e.g., \citealt{Bonazzola1996A&A...312..675B, Owen1998PhRvD..58h4020O,
Miller2019ApJ...887L..24M, Riley2019ApJ...887L..21R}). 

For a constant $n$, we can rewrite Equation~(\ref{eq:sd}) in terms of $P=2\pi/\Omega$ and
$\dot{P}= - 2\pi \dot{\Omega}/\Omega^2$ to describe the spin-down evolution of the pulsars
on the $P$-$\dot{P}$ diagram, i.e., 
\begin{equation}
\frac{{\rm d}P}{{\rm d}t} \propto P^{2-n}.
\end{equation}
The integration of this equation gives the solution 
\begin{equation}\label{eq:eq3}
P_0 =  P \left[1-(n-1) \tau_{\rm p} \dot{P}/P\right]^{\frac{1}{(n-1)}}, \quad \text{for}\ n\neq 1,
\end{equation}
where $\tau_{\rm p}$ is the true age of the pulsar, 
and $P_0$ is its initial period at $\tau_{\rm p}=0$.
We have assumed in Equation~(\ref{eq:eq3}) that $\dot{P}$ is constant over the time-scale of $\tau_{\rm p}$ for a young pulsar. 
In the special case of $n=1$, i.e., where the torque is dominated by a stellar particle wind, the solution reduces to 
\begin{equation}\label{eq:eq4}
P_0 = P\,\exp\left( -\frac{\tau_{\rm p}}{2\tau_{\rm c}}\right),
\end{equation}
with $\tau_{\rm c} = P/2\dot{P}$ being the characteristic age of a pulsar under the assumptions of $P_0 \ll P$ and $n=3$. 

Equations~(\ref{eq:eq3}) and (\ref{eq:eq4}) allow us to estimate the initial period of each pulsar by utilizing their measured values of $P$, $\dot{P}$, and $n$, along with their SNR ages. 
As $P$ and $\dot{P}$ are accurately measured through pulsar timing observations, 
their uncertainties can be neglected in estimating the initial periods. 
The long-term braking index, $n$, has been measured for a few pulsars \citep{Espinoza2017MNRAS.466..147E}. 
For pulsars without a measured braking index, 
we first assume a standard value of $n=3$ and 
will examine the effect of $n\neq 3$ in our discussions. 
Since $\dot{P}$ is small and $n$ is generally measured as $0<n<3$ \citep[see][and references therein]{Espinoza2017MNRAS.466..147E}, the estimated $P_0$ is not significantly affected by the choice of $n$, allowing us to disregard its measurement uncertainty.
Finally, the main source of uncertainty in estimating the initial period comes from $\tau_{\rm snr}$, which has been determined using different methods for different SNRs \citep{Suzuki2021ApJ...914..103S}. 
Some SNRs, e.g., G119.5$+$10.2 and G184.6$-$5.8, have well-recorded ``historical ages" with negligible errors, allowing us to estimate their initial periods with a Dirac $\delta$ distribution.
For SNRs with ages determined within certain ranges, we estimate the initial period as a uniform distribution over the derived lower and upper bounds.

To avoid imaginary or zero periods ($P^2_0\le 0$) when taking $n=3$ in Equation~(\ref{eq:eq3}), $\tau_{\rm snr}<\tau_{\rm c}$ must be satisfied, which is an additional selection criterion for our sample\footnote{There are nine (out of 45) pulsars fall into this condition in the sample used by \cite{Igoshev2022MNRAS.tmp.1595I}, in which case $P_0=2$~ms is set.}. 
There are six pulsars with $n$ measurements in this sample.
In the upper panel of Figure~\ref{MyFig2}, we show 
the derived initial periods of the selected pulsars (blue histogram).

\section{Inferring the initial period distribution}\label{sec:sec3}
\subsection{Method}\label{sec:sec3.1}
With the individual estimates of the initial spin period for $N_{\rm obs}=39$ pulsars, we determine the initial spin period distribution through a hierarchical Bayesian inference \citep{Mandel2019MNRAS.486.1086M,Thrane2019PASA...36...10T}. 
We apply the Bayes' Theorem to infer the posterior distribution of the
population parameter ($\Lambda$) from an ensemble of observations $\{\mathcal{D}\}=\{P, \dot{P}, n, \tau_{\rm snr}\}$, 
\begin{equation}\label{eq:eq5}
p(\Lambda, N | \{\mathcal{D}\}) \propto p(\{\mathcal{D}\}| \Lambda, N) p(\Lambda)p(N),
\end{equation}
where $p(\Lambda)$ and $p(N)$ are the priors for $\Lambda$ and $N$ respectively, with $N$ being the total number of a population over the observation period. 
The inclusion of $N$ in Equation~(\ref{eq:eq5}) allows us to infer the event rate. 
The probability of detecting the $N_{\rm obs}$ pulsars in association with SNRs can be modeled with an inhomogeneous Poisson process.
The first term in the right-hand-side of Equation~(\ref{eq:eq5}) defines a population-level (hierarchical) likelihood \citep{Thrane2019PASA...36...10T}, marginalized over the individual measurements ($\theta$):
\begin{equation}\label{eq:eq6}
\begin{split}
p(\{\mathcal{D}\} | \Lambda, N ) 
 \propto (N\xi(\Lambda))^{N_{\rm obs}} e^{-N\xi(\Lambda)} \prod^{N_{\rm obs}}_{i=1} \frac{1}{N_{{\rm s},i}\xi(\Lambda)} \times \\
 \sum_{j=1}^{N_{{\rm s},i}} \frac{\pi(\theta_{i,j}| \Lambda)}{\pi_{\phi}(\theta_{i,j})}.
\end{split}
\end{equation}
Here, $\pi(\theta| \Lambda)$ is the population model; 
$\pi_{\phi}(\theta)$ is the prior utilized for individual-source Bayesian analysis; $N_{{\rm s}}$ represents the number of discrete samples of individual measurements (drawn from a $\delta$ function or a uniform distribution, see section~\ref{sec:sec2.2}); and $\xi(\Lambda)$ accounts for the fraction of detectable sources in a population, defined with
\begin{equation}\label{eq:eq7}
  \xi(\Lambda) = \int{p}_{\rm det}(\theta) \pi({\theta}| \Lambda)d {\theta}. 
\end{equation}
The probability, ${p}_{\rm det}(\theta)$, denotes the detection probability of observing telescopes. It accounts for the selection bias (see Section~\ref{sec:sec3.2.3}). 

In this work, we restrict our inference to the shape of the population rather than the rate. 
Therefore, we marginalize Equation~(\ref{eq:eq5}) over $N$ by using a log-uniform prior $p(N) \propto 1/N$, which does not affect the inference of $\Lambda$, resulting in 
\begin{equation}\label{eq:marginalposterior}
    {p}(\Lambda | \{\mathcal{D}\}) \propto {p}(\Lambda)\prod^{N_{\rm obs}}_{i=1}  \frac{1}{N_{{\rm s},i}\xi(\Lambda)}\sum_{j=1}^{N_{{\rm s},i}} \frac{\pi(\theta_{i,j}| \Lambda)}{\pi_{\phi}(\theta_{i,j})}.
\end{equation} 
Throughout, we take $\pi_{\phi}(\theta)$ as a uniform prior. 
The Bayes factor, which is the ratio of Bayesian evidence,  
\begin{equation}
    \mathcal{B}  = \frac{\int p(\{\mathcal{D}\}| \Lambda, H_1) p(\Lambda| H_1)d\Lambda}{\int p(\{\mathcal{D}\}| \Lambda, H_0) p(\Lambda| H_0)d\Lambda},
\end{equation}
can be used to arbitrate the preference of one population model over another by the data. 
Here, we assume that the models ($H_0$ as the null-hypothesis model and $H_1$ as the alternative model) in question are equally probable before acquiring any knowledge from the data. 

We utilize the open-source software {\tt BILBY} \citep{Ashton2019ApJS..241...27A} to perform the Bayesian analysis.
The posteriors of free parameters are generated by the sampler using the nested sampling techniques implemented in the python package {\tt DYNESTY} \citep{Speagle2020MNRAS.493.3132S}, whilst estimating the evidence.
We use the Python code {\tt ChainConsumer}\footnote{\url{https://github.com/Samreay/ChainConsumer.git}} to analyze the posterior samples with maximum likelihood statistics by interpolating a Gaussian kernel density function.
The results are quoted at the 1-$\sigma$ credible intervals (i.e., the 68.3\% areas below the maximum posterior density) unless noted otherwise. 

\begin{table*}
\renewcommand{\arraystretch}{1.5}
\centering
\caption{The parameters of each parametric function, their descriptions, priors used in the nested sampling, and posterior credible intervals (1$\sigma$) obtained with different considerations of selection effects. We adopt a uniform prior for each parameter. The prior boundaries of 0.0014~s and 24~s are informed by the measured shortest and longest periods in the ATNF Pulsar Catalogue, respectively.}
\begin{tabular}{lllcccc}
\hline\hline
\multicolumn{1}{l}{\multirow{3}{*}{\text{Symbol}}} &
\multicolumn{1}{l}{\multirow{3}{*}{\text{Description (model)}}} &
\multicolumn{1}{l}{\multirow{3}{*}{\text{Prior}}} &
\multicolumn{4}{c}{\text{Posterior Credible Interval}} \\ \cline{4-7}
& & & no selection & no selection & beaming & beaming \\
& & & ($n=3$) & (random $n$) & (LM88) & (TM98) \\
\hline
$A$~(s)               &          Mean       (GS)       &          (0.0014,  24)    &        $0.017^{+0.030}_{-0.012}$  &                           $0.019^{+0.033}_{-0.015}$  &                           $0.028^{+0.053}_{-0.022}$  &                           $0.035^{+0.065}_{-0.028}$  \\
$B$~(s)               &          Standard   deviation  (GS)       &         (0,    5)       &                          $0.490^{+0.065}_{-0.058}$   &                          $0.490^{+0.074}_{-0.063}$   &                          $0.602^{+0.100}_{-0.091}$   &                          $0.67^{+0.14}_{-0.12}$\\
$D$~(s)               &          Mean       (LOGN)     &          (0.0014,  24)    &        $0.113^{+0.024}_{-0.021}$  &                           $0.115^{+0.023}_{-0.021}$  &                           $0.159^{+0.044}_{-0.034}$  &                           $0.185^{+0.090}_{-0.061}$  \\
$E$                   &          Standard   deviation  (LOGN)     &         (0,    5)       &                          $1.04^{+0.15}_{-0.13}$   &                          $1.05^{+0.15}_{-0.13}$   &                          $1.05^{+0.17}_{-0.13}$   &                          $1.17^{+0.20}_{-0.19}$     \\
$P_{\rm               min}$~(s)  &          Starting   point      (TOPL)    &      (0.007,  0.2)                       &                           $0.008^{+0.002}_{-0.001}$  &                           $0.008^{+0.002}_{-0.001}$  &                           $0.008^{+0.002}_{-0.001}$  &                            $0.008^{+0.002}_{-0.001}$  \\
$F$                   &          Power-law  index      (TOPL)     &         (0,    5)       &                          $1.83^{+0.21}_{-0.18}$   &                          $1.81^{+0.19}_{-0.18}$   &                          $1.49^{+0.20}_{-0.18}$   &                          $1.31^{+0.21}_{-0.19}$    \\
$G$~(s)               &          Width      over       which      $P_0$     turns  over     (TOPL)                     &                           (0,                        0.2)                        &                          $0.057^{+0.016}_{-0.016}$   &                          $0.055^{+0.016}_{-0.017}$    &                          $0.057^{+0.015}_{-0.016}$  &  $0.050^{+0.016}_{-0.018}$  \\
$H$                   &          Shape      parameter  (GAMMA)    &         (0,    10)      &                          $0.82^{+0.19}_{-0.17}$   &                          $0.82^{+0.19}_{-0.18}$   &                          $1.030^{+0.220}_{-0.190}$   &                          $0.95^{+0.20}_{-0.20}$    \\
$I$~(s)               &          Scale      parameter  (GAMMA)    &         (0,    10)      &                          $0.251^{+0.097}_{-0.069}$   &                          $0.255^{+0.100}_{-0.076}$   &                          $0.29^{+0.13}_{-0.11}$   &                          $0.37^{+0.22}_{-0.17}$    \\
$J$                   &          Shape      parameter  (Weibull)  &         (0,    10)      &                          $1.20^{+0.17}_{-0.18}$   &                          $1.16^{+0.17}_{-0.16}$   &                          $1.01^{+0.17}_{-0.17}$   &                          $0.92^{+0.17}_{-0.17}$    \\
$K\, (\mathrm{s}^{-1})$  &          Scale      parameter  (Weibull)  &         (0,    50)      &                          $14.5^{+2.6}_{-2.2}$  &                          $14.0^{+2.7}_{-2.1}$  &                          $10.6^{+2.1}_{-2.1}$  &                          $9.6^{+2.3}_{-2.4}$    \\
\hline
\end{tabular}
\label{MyTabB}
\end{table*}

\begin{figure}
\centering
\includegraphics[width=\columnwidth]{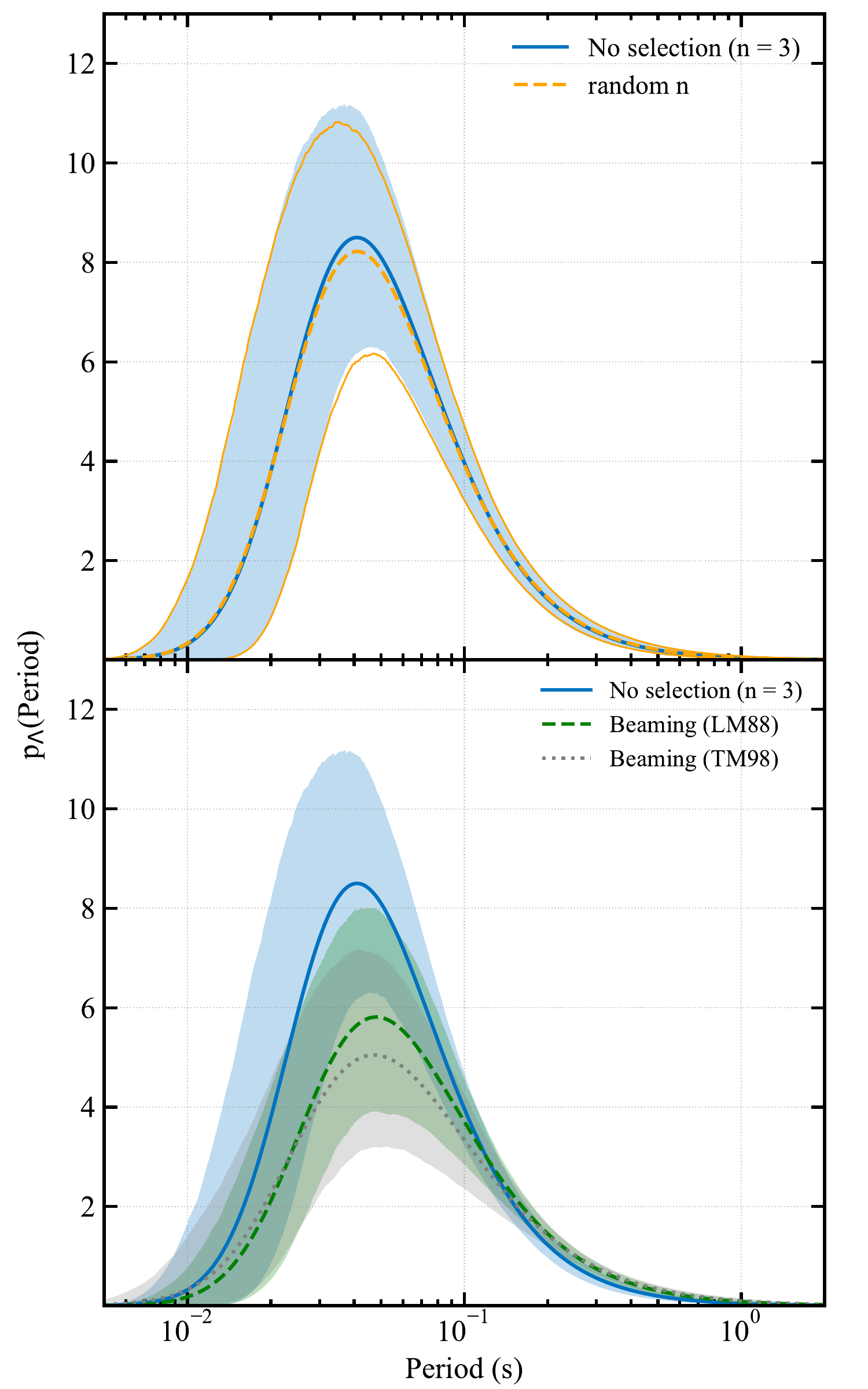}
\caption{The posterior predictive distributions of NS initial spin periods. Upper panel: The Weilbull distribution derived without correcting for selection effects and by adopting fixed (blue) and random (orange) values of braking index. 
Lower panel: The Weilbull distribution derived by correcting the beaming fraction using the LM88 (green) and TM98 (gray) models (see Section \ref{sec:sec3.2.3} for details). The shaded areas indicate the $90\%$ credible intervals. }
\label{MyFigB}
\end{figure}

\subsection{Results and Discussions}\label{sec:sec3.2}
We fit the initial spin period distribution using five parametric functions as population models: the Gaussian (GS), log-normal (LOGN), 
turn-on-power-law (TOPL), GAMMA, and Weibull distributions. 
These functions are described in Appendix~\ref{sec:App2}. 
We explore the posteriors of population parameters using  uniform priors.  
The range of the priors and the definition of each parameter are tabulated in Table~\ref{MyTabB}.

\subsubsection{Without correcting selection effects}\label{sec:sec3.2.1}
We first assume that the observed pulsars are 
representative of the entire population without any selection bias, 
i.e., there is no preference of the detactability for pulsars with different periods. 
Table~\ref{MyTabB} provides the 1$\sigma$ credible intervals 
of marginalized posterior probability distributions of population parameters 
corresponding to each model. 
In Table~\ref{MyTabC}, we present the Bayes factors of different models by taking the GS case as the null hypothesis. 
We obtain decisive evidence for rejecting the GS model. 
This result is in agreement with that obtained by \cite{Igoshev2022MNRAS.tmp.1595I}, 
who found a log-normal distribution provides a better fit to the measurements than the Gaussian based on the Akaike information criterion. 
The Weibull distribution emerges as the most favored, 
being moderately (strongly) preferred over the TOPL (LOGN) 
by a Bayes factor of 7 (276).

\begin{table}
\renewcommand{\arraystretch}{1.5}
\footnotesize
\centering
\caption{The Bayes factors against the Gaussian distribution compared to different models with different considerations of selection effects. }
\begin{tabular}{lcccc}
\hline\hline
\multicolumn{1}{l}{\multirow{2}{*}{\text{Selection effects}}} & \multicolumn{4}{c}{$\ln \mathcal{B}$ \text{relative to the GS distribution}} \\ \cline{2-5}
 & Weibull &   TOPL    &   LOGN    &  GAMMA  \\ \hline
No selection ($n=3$)   &  36.22  &  34.34  &  30.60  &   23.13  \\
No selection (random $n$)   &  34.52  &  32.25  &  29.53  &   22.64  \\
Beaming (LM88)     &        25.60  &  24.56  &  21.05  &  14.07  \\
Beaming (TM98)     &        22.32  &  21.85  &  17.77  &  12.31  \\
\hline
\end{tabular}
\label{MyTabC}
\end{table}

Figure~\ref{MyFig2} plots the posterior predictive distribution (PPD) defined as:
\begin{equation}\label{eq:eq12}
    p_{\Lambda}(\theta'|\{\mathcal{D}\}) = \int  {p}(\Lambda|\{\mathcal{D}\})\pi(\theta'|\Lambda) d\Lambda.
\end{equation}
The PPD of the Weibull and LOGN model are shown in green and grey, respectively. The shaded areas denotes the 90\% credible interval. 
The Weibull distribution, which is the preferred model, indicates that the majority of the observed pulsars were initially fast spinning, 
with initial periods peaking at $\sim$40~ms. 
The 90\% credibility upper limit is constrained to be $P_0^{90\%} \approx 0.5$~s. 
The cumulative distributions of the Weibull and LOGN are shown in the lower panel of Figure~\ref{MyFig2}, where one can see that the Weibull model provides a better fit than the LOGN. 
In Appendix~\ref{sec:AppC}, we compare the initial period distribution inferred with the LOGN model with that obtained by \cite{Igoshev2022MNRAS.tmp.1595I}.

\begin{table}
\small
\centering 
\caption{The results of two-sample Kolmogorov-Smirnoff test for different distribution functions.}
    \begin{tabular}{ccc}
    \hline\hline
    Model & Test statistic & p-value \\
    \hline 
    Weibull & 0.070 & 0.686 \\
    TOPL & 0.133 & 0.053 \\
    LOGN & 0.115 & 0.129 \\
    GAMMA & 0.207 & $3.18\times 10^{-4}$ \\
    GS & 0.437 & $<10^{-4}$ \\
    \hline
    \end{tabular}
\label{MyTab_KS}
\end{table}

To additionally provide a quantitative measure of the choices of the distribution functions, we perform a two-sample Kolmogorov–Smirnov test implemented by the {\tt scipy} package \citep{Virtanen2021zndo...4718897V}\footnote{Two-sample Kolmogorov–Smirnov test evaluates whether two samples are subject to the same underlying distribution or not.}. 
The compared samples are (1) the initial periods from 39 pulsars (as presented in section~\ref{sec:sec2}) and (2) 100 synthetic initial periods drawn from each distribution with the median values of the parameters in Table~\ref{MyTabB}; see the column ``no selection ($n=3$)". 
The values of test statistic, defined as the largest deviation between the two compared cumulative distributions, and p-value are presented in Table~\ref{MyTab_KS}. 
We observe from Table~\ref{MyTab_KS} that the Weibull is the best model; the TOPL and LOGN models come next; and the GAMMA and Gaussian models are disfavoured. 
A Kolmogorov–Smirnov test repeated with 200 synthetic initial periods gives consistent results.
Overall, the Kolmogorov–Smirnov test results support Bayesian model selection results.

\subsubsection{Effects of $n\neq3$}\label{sec:sec3.2.2}
In the analysis above, we have assumed $n=3$ for pulsars without measured brake indices.
However, long-term timing observations of young pulsars 
usually give $n \lesssim 3$ \citep{Espinoza2017MNRAS.466..147E}, 
and $n>3$ are also measured for many glitching or non-glitching pulsars 
\citep{Parthasarathy2019MNRAS.489.3810P, Lower2021MNRAS.508.3251L}.
Such measurements hint at various braking mechanisms or unknown noises in the interior of pulsars during their evolution, which inevitably induce uncertainties in the $n$ measurements.

Here, we explore the dependence of the inferred initial period distribution on 
different values of $n$. 
We randomly draw $n$ values from a Gaussian distribution 
with the mean and standard deviation taken from measured values for pulsars with known $n$. 
For pulsars without a measured braking index, 
we randomly draw $n$ values from a Gaussian distribution 
centered at 3, with a standard deviation of 1, 
constrained to $0\le n \le 4$. 

The resultant posteriors, 
given in Table~\ref{MyTabB}, are consistent with the results in the case of $n=3$. 
The Bayes factors, presented in Table~\ref{MyTabC}, also indicate that 
the adopted values of $n$ does not impact our results significantly. 
From the initial period distribution, shown in Figure~\ref{MyFigB} (in orange), we find that the Weibull distribution is consistent with that obtained in Section~\ref{sec:sec3.2.1} at $90\%$ credibility.  
Therefore, our results are insensitive to the actual values of $n$.

\subsubsection{Correcting for selection effects}\label{sec:sec3.2.3}
The possible selection effects related to radio pulsars, 
SNRs, and CCOs have been discussed in detail by \cite{Igoshev2022MNRAS.tmp.1595I} 
and references therein. 
However, evaluating all the observational effects 
in our population analysis is challenging,
due to different searching strategies and detection methods for different (types of) sources. 
For simplicity, we focus on the known observational selection effects in radio pulsar surveys, such as pulse broadening and the fraction of radio beam directed towards Earth \citep{Lorimer2011ASSP...21...21L}. 
We compute the detection fraction, as defined in Equation~(\ref{eq:eq7}), to correct for these selection effects. 

\begin{figure}
\centering
\includegraphics[width=\columnwidth]{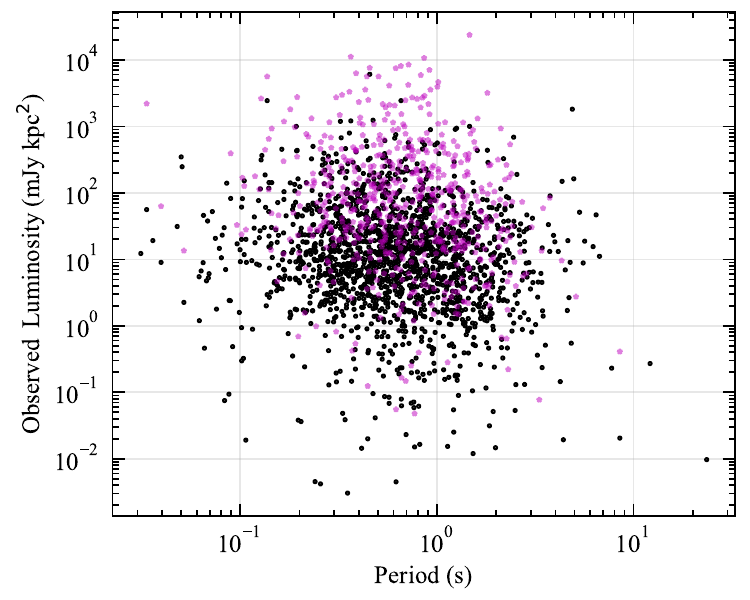}
\caption{The observed radio luminosity versus the spin period. 
The observed radio luminosity is calculated as the product of 
the mean apparent flux density and the square of distance estimated from dispersion measure.
Magenta stars and black dots comprise 583 and 1702 pulsars respectively detected at 400~MHz and 1400~MHz, with $P>10$~ms and $B_{\rm dip}\ge 10^{11}$~G in the ATNF Pulsar Catalogue.}
\label{MyFigC}
\end{figure}

\begin{figure}
\centering
\includegraphics[width=\columnwidth]{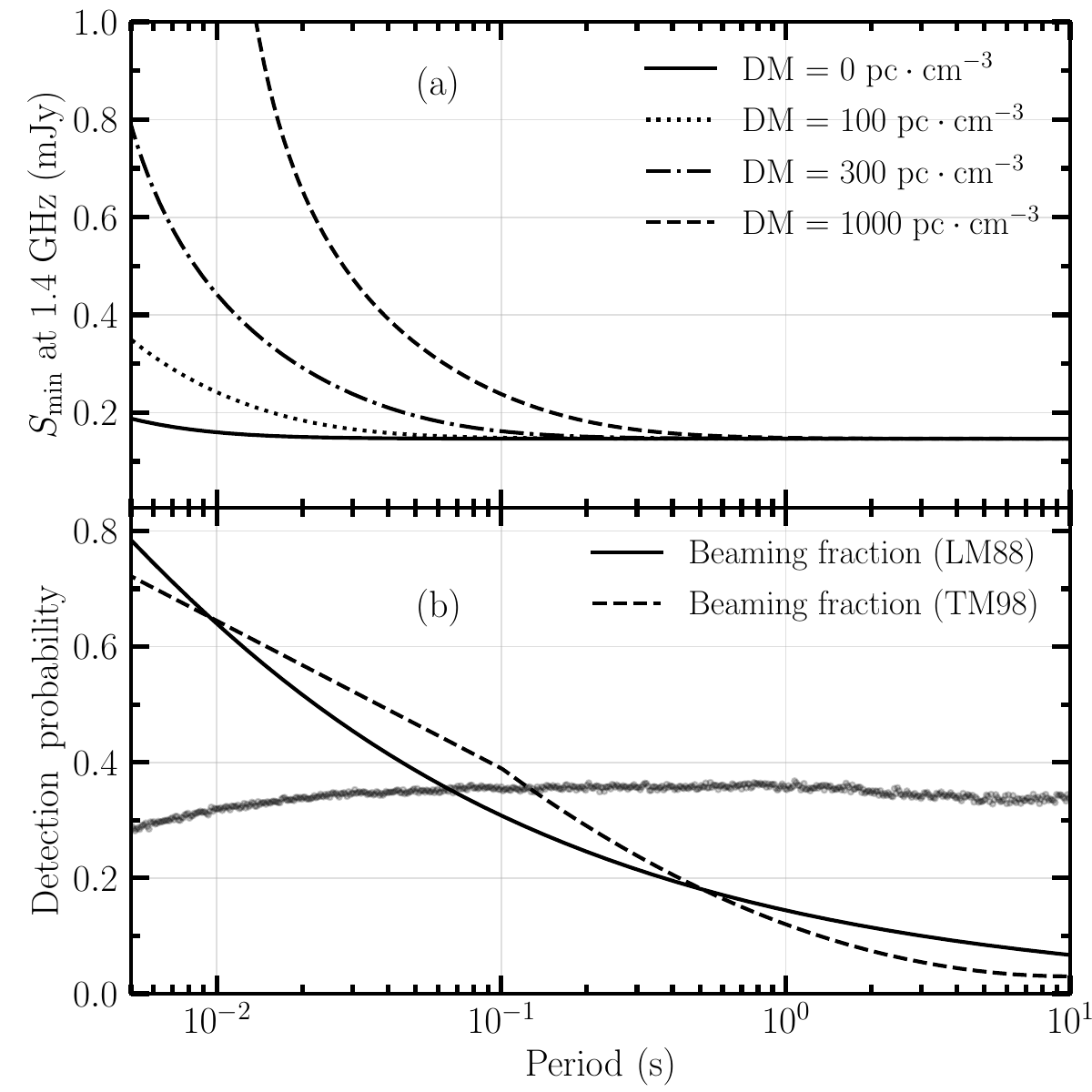}
\caption{Panel (a): The sensitivity of Parkes Multibeam Survey. 
The minimum detectable flux density ($S_{\rm min}$) is computed using Equation~(\ref{eq:fluxlimit}). 
Panel (b): The detectability of radio pulsars. 
The solid and dashed lines account for the beaming fraction of the LM88 and TM98 models, respectively, see Appendix~\ref{sec:App4}. 
The gray dots denote the detectability 
obtained with the inverse of scale factors (see Appendix~\ref{sec:App3}).}
\label{MyFigD}
\end{figure}

The radio selection effects on the spin periods of Galactic pulsar population 
have been extensively studied using the ``current" analysis 
by analytically estimating the pulsar birthrate as a function of period 
\citep{Vivekanand1981JApA....2..315V, Phinney1981MNRAS.194..137P, 
Narayan1987ApJ...319..162N, Narayan1990ApJ...352..222N, 
Lorimer1993MNRAS.263..403L, Vrane2011MNRAS.410.2363V}.
The effects caused by the limited sensitivity in a particular survey 
is typically calculated by the \emph{scale factors} defined as $V_{\rm max}/V$, where $V$ is the weighted volume in which pulsars are detectable and $V_{\rm max}$ is the weighted volume of the whole Galaxy. 
With the pulsar-current analysis, \cite{Vivekanand1981JApA....2..315V} 
proposed an `injection' of a subpopulation of pulsars 
with $\sim$0.5~s 
(see also \citealt{Narayan1987ApJ...319..162N, Narayan1990ApJ...352..222N}). 
However, the injection has been questioned by \cite{Lorimer1993MNRAS.263..403L} 
and \cite{Vrane2011MNRAS.410.2363V}. 
In particular, \cite{Lorimer1993MNRAS.263..403L} pointed out that the injection could be an artifact induced by 
including the luminosity selection effects 
(inverse correlation between luminosity and $P$) 
in the scale factor computations, 
as which would bias the observed sample towards short periods. 

The intrinsic radio-luminosity function remains poorly constrained (see \citealt{Posselt2023MNRAS.520.4582P} and references therein). 
In Figure~\ref{MyFigC}, we show that there is no significant dependence 
of the observed radio (pseudo) luminosity on the period for radio pulsars 
with periods $P>10$~ms and $B_{\rm dip}\ge 10^{11}$~G in the ATNF Catalogue. 
We adopt the observed luminosity distribution to compute the scale factors [see Equation~(\ref{eq:observedLuminosity})]. 
We calculate the scale factors as a function of $P$ 
based on the sensitivity of the Parkes Multibeam Survey\footnote{15 out of 39 in our sample were detected in this survey. The remaining pulsars come from the Fermi Gamma-ray Observatory blind survey, Green Bank Telescope, Arecibo surveys, and other searches.}. 
The sensitivity curve of this survey is shown in panel (a) of Figure~\ref{MyFigD}. 
(The calculation details are described in Appendix~\ref{sec:App3}.) 
Shorter-period and farther pulsars 
are more susceptible to pulse smearing in the interstellar medium, resulting in lower sensitivities in their detection. 
However, the sensitivity becomes almost constant for pulsars with longer periods (e.g., $P\gtrsim 0.1$~s). 
The detection probability, estimated from the inverse of the scale factor, is represented by gray dots in the panel (b). 
The detectability grows slightly with periods for $P\lesssim 0.1$~s, 
which corresponds to the effect of the sensitivity limit on radio observations shown in panel (a).
Overall, the influence of pulse smearing on the shape of the initial period distribution are small, which is consistent with the fact that pulsar luminosity does not depend on spin periods as shown in Figure \ref{MyFigC}.

Pulsars with shorter periods are found to possess wider radio beams, 
resulting in higher detectability in pulsar surveys 
\citep{Lyne1988MNRAS.234..477L, Tauris1998MNRAS.298..625T}. 
Here, we consider two empirical models, LM88 \citep{Lyne1988MNRAS.234..477L} 
and TM98 \citep{Tauris1998MNRAS.298..625T}, 
to quantify the period-beaming fraction correlation (see Appendix~\ref{sec:App4} for details), which is plotted in the panel (b) of Figure~\ref{MyFigD}. 
Incorporating beaming fraction corrections, 
we obtain similar constraints of population parameters 
for both models adopted, see Table~\ref{MyTabB}.  
The Bayes factors given in Table~\ref{MyTabC} indicate that 
the Weibull distribution is still the best description of 
pulsar initial period distribution. 
In the lower panel of Figure~\ref{MyFigB}, one can see a slight shift towards longer periods, 
with a peak at around 50~ms.
The 90\% credibility upper limit is constrained to be $P_0^{90\%}\approx 0.7$~s (0.8~s) for the LM88 (TM98) model.

\section{Conclusions}\label{sec:sec4}
We infer the initial spin period distribution of NSs using 39 pulsars detected in association with SNRs. 
Our hierarchical Bayesian approach accounts for measurement uncertainties 
and known selection effects in radio pulsar surveys.
Using the SNR age as an estimate of the pulsar's true age, 
we determine the initial spin periods of individual pulsars based on their measured values of $P$ and $\dot{P}$, 
assuming a generic spin-down model.
Assuming the observed pulsar sample is unbiased, 
we find that their initial spin period distribution is best described as a Weibull distribution (defined in the spin frequency space), 
peaking at $\sim$40~ms. 
The shape of initial period distribution is not affected by the 
uncertain values of the braking index. 
We also show that the effect of pulse broadening in pulsar surveys 
likely has an insignificant impact on the initial period distribution, 
since the pulsar luminosity appears to be independent of spin periods.
After accounting for selection effects due to pulse beaming fraction, the initial spin periods are still best described as a Weibull distribution, 
with a slight shift towards longer periods and a peak at around $50$~ms.

Our analysis represents a first step in uncovering the birth spin distribution of NSs through rigorous Bayesian population inference.
There are several caveats or assumptions, which may require further investigations. 
First, pulsars are assumed to spin down
with a constant braking index.
Our analysis can be extended to include more complex braking models that allow, e.g., magnetic field decay or decaying magnetic inclination angle.
Second, magnetars are assumed to be a distinct population and thus excluded from our calculations. 
Applying the same analysis to magnetars in SNRs, we obtain initial spin periods $P_0>2$~s, apparently inconsistent with the Weibull distribution found in this work. 
Third, we exclude pulsars whose characteristic ages are lower than SNR ages because in this case no constraint can be placed on their initial spin periods; either these pulsar are born with $\mathcal{O}(\rm{ms})$ spins, or a more complicated braking scenario should be considered, or age estimates of SNRs need to be revisited.
These problems can be at least partly tackled 
by including pulsars not in association with SNRs in the analysis as well as using kinematic or thermal age estimates.

\begin{acknowledgments}
We thank the anonymous referee for valuable comments on
the manuscript. Z.-H. Zhu is supported by the National Natural Science Foundation of China under Grants Nos. 12021003, 11920101003 and 11633001, and the Strategic Priority Research Program of the Chinese Academy of Sciences, Grant No. XDB23000000.
ZQY is supported by the National Natural Science Foundation of China under Grant No. 12305059. XJZ is supported by the National Natural Science Foundation of China (Grant No.~12203004) and by the Fundamental Research Funds for the Central Universities.
ZCC is supported by the National Natural Science Foundation of China (Grant No.~12247176 and No.~12247112) and the innovative research group of Hunan Province under Grant No. 2024JJ1006. 
\end{acknowledgments}

\vspace{5mm}

\software{BILBY \citep{Ashton2019ApJS..241...27A},  
          DYNESTY \citep{Speagle2020MNRAS.493.3132S}, 
          ChainConsumer \citep{Hinton2016, samuel_hinton_2024_10553403}, Scipy \citep{Virtanen2021zndo...4718897V}. 
          }

\clearpage
\appendix

\section{Catalogue of neutron stars associated with SNRs}
\label{sec:append1}

\startlongtable
\begin{deluxetable*}{cllccccccc}
\tablecaption{Parameters of the candidate pulsars observed in association with supernova remnants in the Galaxy. 
References for the $n$ measurements in the table are: (1) \protect\cite{Lyne2015MNRAS.446..857L}, (2) \protect\cite{Espinoza2017MNRAS.466..147E}, (3) \protect\cite{Roy2012MNRAS.424.2213R}, (4) \protect\cite{Gao2016MNRAS.456...55G}, (5) \protect\cite{Weltevrede2011MNRAS.411.1917W}, (6) \protect\cite{Livingstone2011ApJ...742...31L}, (7) \protect\cite{Archibald2016ApJ...819L..16A},  (8) \protect\cite{Archibald2015ApJ...810...67A}. 
\label{MyTabA}}
\small
\tablehead{
\colhead{\#} & \colhead{PSR} & \colhead{SNR} & \colhead{$\tau_{\rm SNR}$} & \colhead{$\tau_{\rm c}$} & \colhead{$P$} & \colhead{$\dot{P}$} & \colhead{$B_{\rm dip}$} & \colhead{$n$} & \colhead{Included}  \\
\colhead{} & \colhead{} & \colhead{} & \colhead{(kyr)} & \colhead{(kyr)} & \colhead{(s)}& \colhead{($\rm s\cdot s^{-1}$)}& \colhead{(G)} & \colhead{} & \colhead{(Y/N)}}
\startdata
1   &  J0002+6216                &  G116.9+00.2     &     7.5$-$18.1   &         306     &    0.115   &      5.97E$-$15  &           8.40E+11  &         -                 &  Y  \\
2   &  J0007+7303                &  G119.5+10.2     &     13$-$13      &         13.9    &    0.316   &      3.60E$-$13  &           1.08E+13  &         -                 &  Y  \\
3   &  J0215+6218                &  G132.7+01.3     &     25$-$33      &         13100   &    0.549   &      6.62E$-$16  &           6.10E+11  &         -                 &  Y  \\
4   &  J0534+2200                &  G184.6$-$5.8 (Crab)    &     0.96$-$0.96  &         1.26    &    0.033   &      4.21E$-$13  &           3.79E+12  &         2.342$\pm$0.001$^{(1)}$   &  Y  \\
5   &  J0538+2817                &  G180.0$-$01.7   &     26$-$34      &         618     &    0.143   &      3.67E$-$15  &           7.33E+11  &         -                 &  Y  \\
6   &  J0821$-$4300$^\star$      &  G260.4$-$03.4   &     2.2$-$5.4    &         193000  &    0.113   &      9.28E$-$18  &           3.27E+10  &         -                 &  Y  \\
7   &  J0835$-$4510              &  G263.9$-$03.3 (Vela)   &     9$-$27       &         11.3    &    0.089   &      1.25E$-$13  &           3.38E+12  &         1.7$\pm$0.2$^{(2)}$       &  Y  \\
8   &  J0855$-$4644              &  G266.2$-$01.2   &     2.4$-$5.1    &         141     &    0.065   &      7.26E$-$15  &           6.94E+11  &         -                 &  Y  \\
9   &  J1016$-$5857              &  G284.3$-$01.8   &     10$-$10      &         21      &    0.107   &      8.08E$-$14  &           2.98E+12  &         -                 &  Y  \\
10  &  J1105$-$6107              &  G290.1$-$00.8   &     10$-$20      &         63.2    &    0.063   &      1.58E$-$14  &           1.01E+12  &         -                 &  Y  \\
11  &  J1157$-$6224              &  G296.8$-$00.3   &     2$-$11       &         1610    &    0.401   &      3.93E$-$15  &           1.27E+12  &         -                 &  Y  \\
12  &  J1210$-$5226$^\star$      &  G296.5+10.0     &     7$-$10       &         302000  &    0.424   &      2.22E$-$17  &           9.83E+10  &         -                 &  Y  \\
13  &  J1322$-$6329              &  G306.3$-$00.9   &     2.5$-$15.3   &         3950    &    2.764   &      1.11E$-$14  &           5.60E+12  &         -                 &  Y  \\
14  &  J1400$-$6325              &  G310.6$-$01.6   &     0.7$-$2      &         12.7    &    0.031   &      3.89E$-$14  &           1.11E+12  &         -                 &  Y  \\
15  &  J1437$-$5959$^{\spadesuit}$              &  G315.9$-$0.0    &     22$-$22      &         114     &    0.062   &      8.59E$-$15  &           7.37E+11  &         -                 &  Y  \\
16  &  J1617$-$5055              &  G332.4$-$00.4   &     2$-$4.4      &         8.13    &    0.069   &      1.35E$-$13  &           3.10E+12  &         -                 &  Y  \\
17  &  J1632$-$4757              &  G336.4+00.2     &     10$-$10      &         240     &    0.229   &      1.51E$-$14  &           1.88E+12  &         -                 &  Y  \\
18  &  J1702$-$4128              &  G344.7$-$00.1   &     3$-$6        &         55.1    &    0.182   &      5.23E$-$14  &           3.12E+12  &         -                 &  Y  \\
19  &  J1721$-$3532              &  G351.7+00.8     &     0$-$68       &         176     &    0.28    &      2.52E$-$14  &           2.69E+12  &         -                 &  Y  \\
20  &  J1747$-$2809              &  G000.9+00.1     &     1.9$-$1.9    &         5.31    &    0.052   &      1.56E$-$13  &           2.88E+12  &         -                 &  Y  \\
21  &  J1747$-$2958              &  G359.1$-$00.5   &     17$-$20.7    &         25.5    &    0.099   &      6.13E$-$14  &           2.49E+12  &         -                 &  Y  \\
22  &  J1801$-$2304              &  G006.4$-$00.1   &     33$-$36      &         58.3    &    0.416   &      1.13E$-$13  &           6.93E+12  &         -                 &  Y  \\
23  &  J1801$-$2451$^{\spadesuit}$              &  G5.4$-$1.2      &     4.7$-$5.2    &         15.5    &    0.125   &      1.28E$-$13  &           4.04E+12  &         1.1$\pm$0.4$^{(2)}$       &  Y  \\
24  &  J1803$-$2137              &  G008.7$-$00.1   &     15$-$28      &         15.8    &    0.137   &      1.34E$-$13  &           4.29E+12  &         1.9$\pm$0.5$^{(2)}$       &  Y  \\
25  &  J1809$-$2332              &  G007.5$-$01.7   &     50$-$50      &         67.6    &    0.147   &      3.44E$-$14  &           2.27E+12  &         -                 &  Y  \\
26  &  J1811$-$1925              &  G011.2$-$00.3   &     1.4$-$2.4    &         23.3    &    0.065   &      4.40E$-$14  &           1.71E+12  &         -                 &  Y  \\
27  &  J1813$-$1749              &  G012.8$-$00.0   &     1.2$-$1.2    &         5.58    &    0.045   &      1.27E$-$13  &           2.41E+12  &         -                 &  Y  \\
28  &  J1833$-$1034              &  G021.5$-$00.9   &     1.55$-$1.8   &         4.85    &    0.062   &      2.02E$-$13  &           3.58E+12  &         1.8569$\pm$0.0006$^{(3)}$  &  Y  \\
29  &  J1852+0040$^\star$        &  G033.6+00.1     &     4.4$-$6.7    &         192000  &    0.105   &      8.68E$-$18  &           3.05E+10  &         -                 &  Y  \\
30  &  J1853$-$0004              &  G032.8$-$00.1   &     5.7$-$22     &         288     &    0.101   &      5.57E$-$15  &           7.61E+11  &         -                 &  Y  \\
31  &  J1856+0113                &  G034.7$-$00.4   &     7.9$-$8.9    &         20.3    &    0.267   &      2.08E$-$13  &           7.55E+12  &         -                 &  Y  \\
32  &  J1857+0143                &  G035.6$-$00.4   &     2.3$-$2.3    &         71      &    0.14    &      3.12E$-$14  &           2.11E+12  &         -                 &  Y  \\
33  &  J1906+0722                &  G041.1$-$00.3   &     1.35$-$5.3   &         49.2    &    0.112   &      3.59E$-$14  &           2.02E+12  &         -                 &  Y  \\
34  &  J1952+3252                &  G069.0+02.7     &     60$-$60      &         107     &    0.04    &      5.84E$-$15  &           4.86E+11  &         -                 &  Y  \\
35  &  J1957+2831                &  G065.1+00.6     &     40$-$140     &         1570    &    0.308   &      3.11E$-$15  &           9.90E+11  &         -                 &  Y  \\
36  &  J2021+4026                &  G078.2+02.1     &     8$-$16       &         76.9    &    0.265   &      5.47E$-$14  &           3.85E+12  &         -                 &  Y  \\
37  &  J2047+5029                &  G089.0+04.7     &     4.8$-$18     &         1690    &    0.446   &      4.18E$-$15  &           1.38E+12  &         -                 &  Y  \\
38  &  J2229+6114                &  G106.3+02.7     &     3.9$-$12     &         10.5    &    0.052   &      7.83E$-$14  &           2.03E+12  &         0$\pm$1$^{(2)}$           &  Y  \\
39  &  J2337+6151                &  G114.3+00.3     &     7.7$-$7.7    &         40.6    &    0.495   &      1.93E$-$13  &           9.91E+12  &         -                 &  Y  \\
\\
40  &  J0205+6449                &  G130.7+03.1     &     4.3$-$7      &         5.37    &    0.066   &      1.94E$-$13  &           3.61E+12  &         -                 &  N  \\
41  &  J0501+4516$^{\dagger}$    &  G160.9+02.6     &     2.6$-$9.2    &         15.7    &    5.762   &      5.82E$-$12  &           1.85E+14  &         6.3$\pm$1.7$^{(4)}$       &  N  \\
42  &  J0502+4654                &  G160.9+02.6     &     2.6$-$9.2    &         1810    &    0.639   &      5.58E$-$15  &           1.91E+12  &         -                 &  N  \\
43  &  J0554+3107$^{\spadesuit}$                &  G179.0+2.6      &     -            &         51.7    &    0.465   &      1.43E$-$13  &           8.24E+12  &         -                 &  N  \\
44  &  J0630$-$2834              &  G276.5+19.0     &     1000$-$6000  &         2770    &    1.244   &      7.12E$-$15  &           3.01E+12  &         -                 &  N  \\
45  &  J0633+0632$^{\spadesuit}$                &  G205.5+0.5      &     30$-$150     &         59.2    &    0.297   &      7.96E$-$14  &           4.92E+12  &         -                 &  N  \\
46  &  J0659+1414$^{\spadesuit}$                &  Monogem         Ring  &            86$-$170  &       111  &       0.385  &           5.49E$-$14  &         4.65E+12  &                 -  &  N   \\
47  &  J0953+0755                &  G276.5+19.0     &     1000$-$6000  &         17500   &    0.253   &      2.30E$-$16  &           2.44E+11  &         -                 &  N  \\
48  &  J1101$-$6101              &  G290.1$-$00.8   &     10$-$20      &         116     &    0.063   &      8.56E$-$15  &           7.42E+11  &         -                 &  N  \\
49  &  J1119$-$6127              &  G292.2$-$00.5   &     4.2$-$7.1    &         1.61    &    0.408   &      4.02E$-$12  &           4.10E+13  &         2.684$\pm$0.002$^{(5)}$   &  N  \\
50  &  J1124$-$5916              &  G292.0+01.8     &     2.93$-$3.05  &         2.85    &    0.135   &      7.53E$-$13  &           1.02E+13  &         -                 &  N  \\
51  &  J1341$-$6220$^{\spadesuit}$              &  G308.8$-$0.1    &     0$-$32.5     &         12.1    &    0.193   &      2.53E$-$13  &           7.08E+12  &         -                 &  N  \\
52  &  J1513$-$5908              &  G320.4$-$01.2   &     1.9$-$1.9    &         1.57    &    0.152   &      1.53E$-$12  &           1.54E+13  &         2.832$\pm$0.003$^{(6)}$   &  N  \\
53  &  J1522$-$5735$^{\spadesuit}$              &  G321.9$-$0.3    &     -            &         51.8    &    0.102   &      3.12E$-$14  &           1.81E+12  &         -                 &  N  \\
54  &  J1550$-$5418$^{\dagger}$$^{\spadesuit}$  &  G327.24$-$0.13  &     -            &         1.41    &    2.07    &      2.32E$-$11  &           2.22E+14  &         -                 &  N  \\
55  &  J1614$-$5048              &  G332.4+00.1     &     3$-$8.6      &         7.42    &    0.232   &      4.95E$-$13  &           1.08E+13  &         -                 &  N  \\
56  &  J1622$-$4944              &  G333.9+00.0     &     0$-$6        &         995     &    1.073   &      1.71E$-$14  &           4.33E+12  &         -                 &  N  \\
57  &  J1622$-$4950$^{\dagger}$  &  G333.9+00.0     &     0$-$6        &         24.7    &    4.327   &      2.78E$-$12  &           1.11E+14  &         -                 &  N  \\
58  &  J1632$-$4818$^{\spadesuit}$              &  G336.1$-$0.2    &     -            &         19.9    &    0.814   &      6.49E$-$13  &           2.33E+13  &         -                 &  N  \\
59  &  J1635$-$4735$^{\dagger}$  &  G337.0$-$00.1   &     5$-$5        &         -       &    2.595   &      -           &           -         &         -                 &  N  \\
60  &  J1640$-$4631              &  G338.3$-$00.0   &     1$-$8        &         3.35    &    0.206   &      9.76E$-$13  &           1.44E+13  &         3.15$\pm$0.03$^{(7)}$     &  N  \\
61  &  J1640$-$4631              &  G338.5$-$00.1   &     1.1$-$17.0        &         3.35    &    0.206   &      9.76E$-$13  &           1.44E+13  &         3.15$\pm$0.03$^{(7)}$     &  N  \\
62  &  J1646$-$4346$^{\spadesuit}$              &  G341.2+0.9      &     -            &         32.5    &    0.232   &      1.13E$-$13  &           5.17E+12  &         -                 &  N  \\
63  &  J1709$-$4429$^{\spadesuit}$              &  G343.1$-$2.3    &     -            &         17.5    &    0.102   &      9.30E$-$14  &           3.12E+12  &         -                 &  N  \\
64  &  J1714$-$3810$^{\dagger}$  &  G348.7+00.3     &     0.65$-$16.8  &         1.03    &    3.825   &      5.88E$-$11  &           4.80E+14  &         -                 &  N  \\
65  &  J1726$-$3530$^{\spadesuit}$              &  G352.2$-$0.1    &     -            &         14.5    &    1.11    &      1.22E$-$12  &           3.72E+13  &         -                 &  N  \\
66  &  J1731$-$4744$^{\spadesuit}$              &  G343.0$-$6.0    &     -            &         80.4    &    0.83    &      1.64E$-$13  &           1.18E+13  &         -                 &  N  \\
67  &  J1808$-$2024$^{\dagger}$$^{\spadesuit}$  &  G10.0$-$0.3     &     -            &         0.218   &    7.556   &      5.49E$-$10  &           2.06E+15  &         -                 &  N  \\
68  &  J1809$-$1917              &  G011.2$-$00.3   &     1.4$-$2.4    &         51.4    &    0.083   &      2.55E$-$14  &           1.47E+12  &         -                 &  N  \\
69  &  J1809$-$1943$^{\dagger}$  &  G011.2$-$00.3   &     1.4$-$2.4    &         31      &    5.541   &      2.83E$-$12  &           1.27E+14  &         -                 &  N  \\
70  &  J1833$-$0827              &  G023.3$-$00.3   &     60$-$200     &         147     &    0.085   &      9.18E$-$15  &           8.95E+11  &         -                 &  N  \\
71  &  J1834$-$0845$^{\dagger}$  &  G023.3$-$00.3   &     60$-$200     &         4.94    &    2.482   &      7.96E$-$12  &           1.42E+14  &         -                 &  N  \\
72  &  J1841$-$0456$^{\dagger}$  &  G027.4+00.0     &     0.75$-$2.1   &         4.57    &    11.789  &      4.09E$-$11  &           7.03E+14  &         -                 &  N  \\
73  &  J1844$-$03$^{\spadesuit}$                &  G29.6+0.1       &     0$-$8        &         -       &    6.971   &      -           &           -         &         -                 &  N  \\
74  &  J1846$-$0258              &  G029.7$-$00.3   &     1.69$-$1.85  &         0.728   &    0.327   &      7.11E$-$12  &           4.88E+13  &         2.19$\pm$0.03$^{(8)}$     &  N  \\
75  &  J1850$-$0006$^{\spadesuit}$              &  G32.45+0.1      &     -            &         8040    &    2.191   &      4.32E$-$15  &           3.11E+12  &         -                 &  N  \\
76  &  J1852+0033$^{\dagger}$    &  G033.6+00.1     &     4.4$-$6.7    &         -       &    11.559  &      -           &           -         &         -                 &  N  \\
77  &  J1857+0210                &  G035.6$-$00.4   &     2.3$-$2.3    &         712     &    0.631   &      1.40E$-$14  &           3.01E+12  &         -                 &  N  \\
78  &  J1857+0212                &  G035.6$-$00.4   &     2.3$-$2.3    &         164     &    0.416   &      4.03E$-$14  &           4.14E+12  &         -                 &  N  \\
79  &  J1907+0602$^{\spadesuit}$                &  G40.5$-$0.5     &     -            &         19.5    &    0.107   &      8.68E$-$14  &           3.08E+12  &         -                 &  N  \\
80  &  J1907+0631$^{\spadesuit}$                &  G40.5$-$0.5     &     20$-$40      &         11.3    &    0.324   &      4.52E$-$13  &           1.22E+13  &         -                 &  N  \\
81  &  J1907+0919$^{\dagger}$$^{\spadesuit}$    &  G42.8+0.6       &     -            &         0.895   &    5.198   &      9.20E$-$11  &           7.00E+14  &         -                 &  N  \\
82  &  J1913+1011                &  G044.5$-$00.2   &     70$-$200     &         169     &    0.036   &      3.37E$-$15  &           3.52E+11  &         -                 &  N  \\
83  &  J1930+1852                &  G053.4+0.00     &     2$-$5        &         2.89    &    0.137   &      7.51E$-$13  &           1.03E+13  &         -                 &  N  \\
84  &  J1930+1852                &  G054.1+00.3     &     2$-$5        &         2.89    &    0.137   &      7.51E$-$13  &           1.03E+13  &         -                 &  N  \\
85  &  J1932+1916                &  G054.4$-$00.3   &     61$-$61      &         35.4    &    0.208   &      9.32E$-$14  &           4.46E+12  &         -                 &  N  \\
86  &  J1935+2154$^{\dagger}$    &  G057.2+00.8     &     16$-$95      &         -       &    3.245   &      -           &           -         &         -                 &  N  \\
87  &  J2022+3842$^{\spadesuit}$                &  G76.9+1.0       &     -            &         8.94    &    0.049   &      8.61E$-$14  &           2.07E+12  &         -                 &  N  \\
88  &  J2301+5852$^{\dagger}$    &  G109.1$-$01.0   &     8.8$-$14     &         235     &    6.979   &      4.71E$-$13  &           5.80E+13  &         -                 &  N  \\
\enddata
\tablenotetext{\dagger}{The objects known as magnetars in the McGill Online Magnetar Catalog.}
\tablenotetext{\star} {The objects known as central compact objects.}
\tablenotetext{\spadesuit}{The objects are not included in the sample of \cite{Igoshev2022MNRAS.tmp.1595I}.}
\end{deluxetable*}

\section{Parametric functions for the spin period distribution}\label{sec:App2}

(I) Gaussian distribution: 

\begin{equation}\label{eq:gauss}
\pi(P_0 |  A, B )
=\frac{C}{\sqrt{2 \pi} B} \exp \left[-\frac{1}{2}\left(\frac{P_0 -A}{ B}\right)^{2}\right],
\end{equation}
where $A$ and $B$ are the mean and the standard deviation, respectively, and $C$ is the normalization factor:

\begin{equation}
      C = \left(1- \frac{1}{2}\left[ 1+{\rm erf}\left(-\frac{A}{\sqrt{2} B} \right)\right]\right)^{-1}.
\end{equation}

(II) The log-normal distribution:
\begin{equation}
\pi(P_0 | D, E) = \frac{1}{\sqrt{2 \pi} E P_0} \exp \left[-\frac{\ln^2 (P_0/D)}{2E^2}\right],
\end{equation}
where $D$ and $E$ are the mean and the standard deviation, respectively. 

(III) The turn-on-power-law (TOPL) distribution \citep{Talbot2018ApJ...856..173T}:
\begin{equation}\label{eq:top}
    \pi(P_0 | P_{\rm min}, F, G) \propto P_0^{-F} S(P_0, P_{\rm min},  G).
\end{equation}
Here, $F$ is the power-law index of the TOPL model and $G$ is the width over which $P_0$ turns on.  
The smoothing function $S$ applies to the part starting from a minimum value, rising from 0 at $P_{\rm min}$ to unity at $P_{\rm min} + G$, i.e., 
\begin{equation}\label{eq:smooth}
    S(P_0, P_{\rm min }, G)=(\exp [f(P_0-P_{\rm min }, G)]+1)^{-1},
\end{equation}
\begin{equation}
    f(P_0, G) =\frac{G}{P_0}-\frac{G}{P_0-G }.
\end{equation}
For $G = 0$, Equation~(\ref{eq:smooth}) returns to a step function, and the TOPL model returns to a simple power-law function.
This distribution is normalized by numerically integrating Equation~(\ref{eq:top}) over [$P_{\rm min}$, $\infty$). 

(IV) The GAMMA distribution: 
\begin{equation}
\pi(P_0 | H, I) =  \frac{P_0^{H-1}}{I^H \Gamma(H)}\exp{\left(-\frac{P_0}{I}\right)}, 
\end{equation}
where $H$ and $I$ denote the shape and scale parameters of the distribution, respectively.

(V) The Weibull distribution \citep{PHA17, LYC+23} transformed to the period space:
\begin{equation}\label{eq:weibull}
\pi_{\nu}(P_0 | J, K) = \frac{J}{KP^2_0}\left(\frac{1}{K P_0}\right)^{J-1}\exp\left[-\left(\frac{1}{P_0K}\right)^J\right],
\end{equation}
where $J$ and $K$ are known as the shape and scale parameters of the distribution, respectively. 
The subscript `$\nu$' means that the corresponding distribution in frequency space is Weibull. 
The Weibull distribution peaks at $K^{-1}[J/(J+1)]^{1/J}$.

\section{Comparison with previous works}\label{sec:AppC}

\begin{figure}
\centering
\includegraphics[width=\columnwidth]{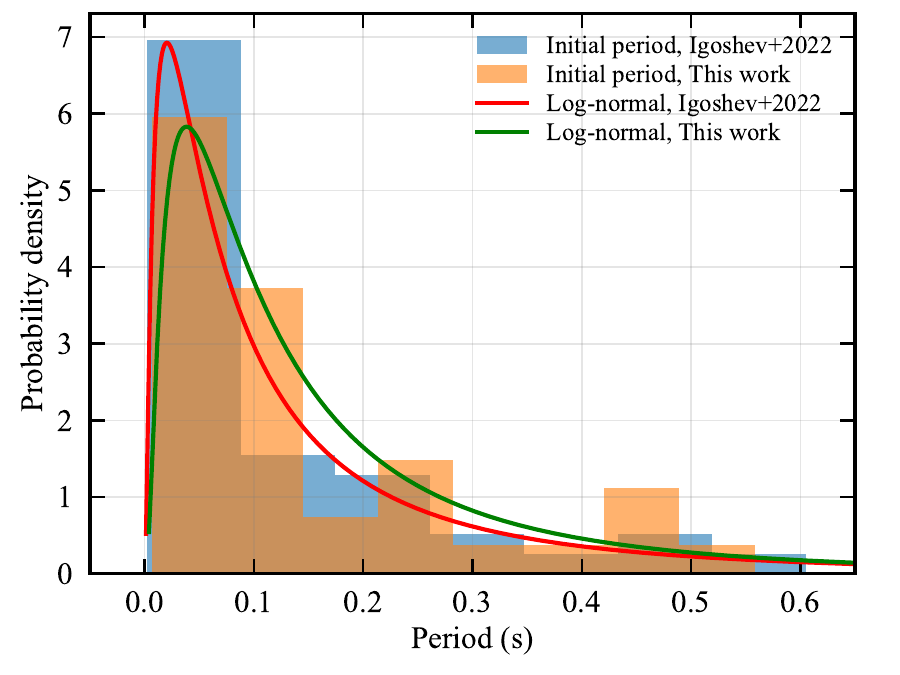}
\caption{The log-normal distribution obtained in this work (green) and that in \cite{Igoshev2022MNRAS.tmp.1595I} (red). The histograms show the initial periods used by \cite{Igoshev2022MNRAS.tmp.1595I} (in blue) along with that analyzed in this work (in orange). 
}
\label{MyFig_compare}
\end{figure}

Figure~\ref{MyFig_compare} displays the initial periods data along with the log-normal distributions obtained by \cite{Igoshev2022MNRAS.tmp.1595I} (in red) and inferred in this work (in green); both without the correction of selection effects.
One can see that both sets of results are generally consistent with each other.
However, in our work the log-normal model is disfavored. This is due to a different analysis method, and more importantly different data selection criteria, especially the treatment of the observed pulsars with supposedly imaginary initial periods (see Section~\ref{sec:sec2.2}).

\section{Scale factor computations}\label{sec:App3}
The scale factor for a pulsar with period, $P$, is defined as
\begin{equation}\label{eq:scalefactor}
     \psi(P) = \frac{\iiint \rho_{\rm r}(r)\rho_{\rm z}(z) r dr d\phi dz}{\iiiint' \rho_{\rm r}(r)\rho_{\rm z}(z) \eta(P, L, r, \phi, z) r d r d\phi dzdL}.
\end{equation}
Here, $\rho_{\rm r}$ represents the space density for the galactocentric radius ($r$), 
and $\rho_{\rm z}$ is the density in the height above the Galactic plane ($z$). 
We assume that the pulsars distribute uniformly over the galactocentric azimuthal angle ($\phi$), 
which is justified since there is no apparent concentration of pulsars 
observed in the inner spiral arms of the Galaxy. 
The selection effects caused by the telescope's limited sensitivity 
enter in the parameter $\eta(P, L, r, \phi, z)$ via the inverse-square law, $S_{\rm \nu}=L d^{-2}$, 
where $S_{\rm \nu}$ is the apparent flux density detectable at a center frequency 
$\nu_{\rm center}$ (throughout we take $\nu_{\rm center}=1.4$~GHz). 
Given a minimum detectable flux density, $S_{\rm min}$, in a specific survey, 
$\eta(P, L, r, \phi, z)$ is set to 1 if $S_{\nu}\ge S_{\rm min}$ for a pulsar with period $P$ and luminosity $L$ 
at coordinates $\{r, \phi, z\}$, or otherwise it is set to 0.
The numerator in Equation~(\ref{eq:scalefactor}) is integrated over 
the whole volume of the Galaxy, while the denominator is only integrated over the volumes 
where pulsars can be potentially detectable in any reference surveys. 

The integral in Equation~(\ref{eq:scalefactor}) can be numerically computed at a number of values of $P$ utilizing Monte Carlo simulations. 
For each pulsar with $P$, $S_{\nu}$ and $S_{\rm min}$ are computed at a large number of grid points in the
$\{r, \phi, z, L\}$ space. 
Then the numerator, which can be seen as the \emph{wighted} total volume, is updated by adding the weight $\rho(r)\rho(z)$.
The denominator, which can be thought to be the \emph{weighted} sensitive volume, is updated by adding the weight $\rho(r)\rho(z)$ if $S_{\nu}\ge S_{\rm min}$. 
The settings are given below. 

{\bf Spatial distribution.} 
We use the $r$-distribution \citep{Yusifov2004A&A...422..545Y}
\begin{equation}
\rho(r) = 37.6\left(\frac{r+0.55}{R_{\odot}+0.55}\right)^{1.64} \;\exp{\left[ -4\left( \frac{r-R_{\odot}}{R_{\odot}+0.55}\right)\right]},
\end{equation}
where $R_{\odot}=8.5$~kpc is the galactocentric distance of the Sun. 
For the $z$-distribution, we use the exponential function \citep{Lorimer2006MNRAS.372..777L}
\begin{equation}
\rho(z) = 0.75\exp{\left(-\frac{\lvert z\lvert}{0.18~\rm kpc} \right)}.
\end{equation}

\begin{figure}
   \centering
   \includegraphics[angle=0,scale=0.5]{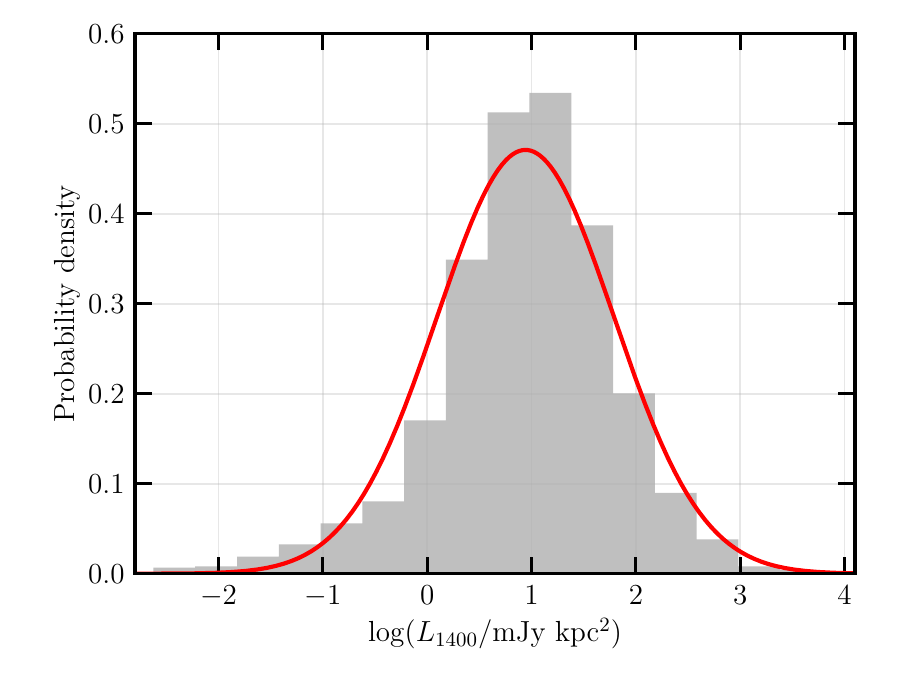}
   \caption{The distribution of the observed radio luminosities of 1834 pulsars with $P>10$~ms detected at 1400~MHz 
in the ATNF Catalogue. The red line is obtained by fitting the data to a log-normal distribution (see Equation~\ref{eq:observedLuminosity}).
   }
    \label{MyFigE}
\end{figure}

{\bf Luminosity distribution.} 
We produce the luminosities of model pulsars with the observed luminosity distribution (see Figure~\ref{MyFigE}). 
We fit the observed radio luminosities (calculated as the product of 
the mean apparent flux density at 1400~MHz and the square of distance) of 1834 pulsars ($P>10$~ms) with a log-normal distribution to obtain 
\begin{equation}\label{eq:observedLuminosity}
p(\log(L_{1400})) \propto \exp\left[-\frac{1}{2} \left(\frac{\log(L_{1400}) - 0.94}{0.85}\right)^2\right].
\end{equation}

{\bf Dispersion measure at each position.} At each position in the Galaxy, we compute the dispersion measure (DM) by using the YMW16 free electron density model \citep{Yao2017ApJ...835...29Y}\footnote{\url{https://github.com/FRBs/pygedm}}
\begin{equation}
   {\rm DM} = \int^d_0 n_e {\rm d}l~\rm cm^{-3}\;pc,
\end{equation} 
where $d$ is in units of pc and $n_e$ denotes the free electron density in cm$^{-3}$. 
To account for the errors in estimating the model dispersion measure (DM$_{\rm mod}$) by using this electron density model, we take a random error drawn from a Gaussian distribution with a mean of zero and width of $0.2\,\rm DM_{\rm mod}$.

{\bf Instrumental sensitivity and pulse broadening.} 
The sensitivity limits of radio telescopes are determined by the various instrumental parameters,
including the system noise temperature ($T_{\rm sys}$), antenna gain $G$, number of polarizations $N_{\rm p}$, total bandwidth $\Delta \nu$, and integration time $t_{\rm int}$.
For a specific survey, the theoretical minimum mean flux density can be characterized by the radiometer equation \citep{Dewey1985ApJ...294L..25D}:
\begin{equation}\label{eq:fluxlimit}
S_{\rm min}(P) = \frac{\rho_{\rm s/n} \beta (T_{\rm r}+T_{\rm sky})}{G\sqrt{N_{\rm p} \Delta \nu t_{\rm int}}}\sqrt{\frac{W_{\rm e}}{P-W_{\rm e}}}, \,P>W_{\rm e},
\end{equation}
where $\rho_{\rm s/n}=8$ is the signal-to-noise ratio for the detection threshold, and $\beta \approx 1.5$, which includes the loss from one bit digitization ($\sim \sqrt{\pi/2}$) and other kinds of system losses. 
We scale the sky temperature measured by \cite{Haslam1982A&AS...47....1H} at 408-MHz to any frequency using the sky background spectrum:
\begin{equation}
\frac{T_{\rm sky}(l,b, \nu)}{\rm K} = \left(25 + \frac{275}{[1+(l/42)^2][1+(b/3)^2]}\right) \;\left(\frac{\nu}{408}\right)^{-2.6},
\end{equation}
where the spectral index $-2.6$ is taken from \cite{Lawson1987MNRAS.225..307L}.
The sensitivity will drop when the radio photons from the pulsar are located offset about the beam center, which can be accounted for by using a Gaussian pattern to provide
\begin{equation}
G_{\rm beam}(\mathcal{R}) = G_0\;\exp\left( \frac{-2.77\mathcal{R}^2}{w^2}\right).
\end{equation}
Here, $G_0$ is the antenna gain at the beam center, $w$ is the full width at half maximum of the telescope beam (in arcmin), and $\mathcal{R}$ represents the offset from the beam center, randomly drawn from a Gaussian distribution with a mean of zero and standard deviation of $w/2$.

The observed pulse width incorporating the effects of instrumental settings, dispersion, and scattering can be modeled as \citep{Dewey1985ApJ...294L..25D}
\begin{equation}
    W_{\rm e} = \sqrt{W^2_{\rm int}+ \tau^2_{\rm samp} + \tau^2_{\rm scatt} +  \tau^2_{\rm DM} + \tau^2_{\rm \Delta DM}}.
\end{equation}
The first term is the intrinsic pulse width (in second), which is adopted as $W_{\rm int}=0.04 P$. 
The second term, $\tau_{\rm samp}$, is the data sampling effects, which account for the details of time resolution of digitized data from the system hardware like antialiasing filters.
We take $\tau_{\rm samp}$ as the sampling interval ($t_{\rm samp}$) of the radio telescope.
The third term, $\tau_{\rm scatt}$, is the scatter-broadening time, which  arises from the smearing due to multi-path propagation of light in a non-uniform and ionized interstellar medium, given by \citep{Bhat2004ApJ...605..759B}
\begin{equation}
\begin{split}
\log\left(\frac{\tau_{\rm scatt}}{\rm ms}\right) = & -6.46 + 0.154\log{\rm DM} \\ 
&+ 1.07(\log {\rm DM})^2 - 3.86\log\left(\frac{\nu_{\rm center}}{\rm GHz} \right).
\end{split}
\end{equation}
The fourth term, $\tau_{\rm DM}$, is the dispersion-broadening time across an individual channel, which is adopted as \citep{Hessels2007ApJ...670..363H}
\begin{equation}
\tau_{\rm DM} = 8.3\rm \left(\frac{DM}{pc\;cm^{-3}}\right)\;\left(\frac{\Delta{\nu_{\rm chan}}}{MHz}\right) \;\left(\frac{\nu_{\rm center}}{GHz}\right)^{-3}~\mu s,
\end{equation}
where the channel bandwidth $\Delta{\nu_{\rm chan}}\ll \nu_{\rm center}$ is assumed; for the 1.4~GHz surveys, we employ $\Delta{\nu_{\rm chan}}=100/256$~MHz for $\rm DM<100$~cm$ ^{-3}$~pc and $\Delta{\nu_{\rm chan}}=100/512$~MHz for $\rm DM>100$~cm$ ^{-3}$~pc.
The last term, $\tau_{\Delta \rm DM}$, is the smearing due to the finite DM step size in the survey, which is neglected here.

{\bf Pulsar survey.} We use the sensitivity limit of the Parkes Multibeam Survey  \citep{Manchester2001MNRAS.328...17M} to evaluate the scale factors, its parameters are shown in Table~\ref{MyTabD}. 
The panel (a) of Figure~\ref{MyFigC} displays the sensitivity curve of this survey. 
Note that, to obtain the $S_{\rm min}$ in this plot, we have simply adopted a sky temperature of 25~K, the observed pulse width is approximated as $[(0.04 P)^2 + t^2_{\rm samp} + \tau^2_{\rm DM}]^{1/2}$, and $\tau_{\rm DM}\approx t_{\rm samp}\frac{\rm DM}{\rm DM_0}$ with $\rm DM_0= 28$~cm$^{-3}$~pc.

\section{Beaming fraction}\label{sec:App4}
The empirical model, LM88, is given by \citep{Lyne1988MNRAS.234..477L} 
\begin{equation}
    f_{\rm LM88}(P) = \frac{4}{\pi} \sin (6.5^\circ P^{-1/3}).
\end{equation}

An alternative model, TM98, is given by \citep{Tauris1998MNRAS.298..625T}
\begin{equation}\label{eq:beam1}
f_{\rm TM98}(P>0.1~{\rm s}) \approx 0.09~[\log\,(P/{\rm s})-1]^2+0.03.
\end{equation}
While this relation was derived based on the radio pulsars with luminosities $>0.1$~mJy\;kpc$^2$ and might differ for dimmer objects, we assume it is valid for the dimmer pulsars. 
We extend Equation~(\ref{eq:beam1}) linearly to estimate the beaming fraction 
for pulsars with periods $P \le 0.1$~s: 
with a lower limit of $f_{\rm TM98}(P = 0.1~{\rm s}) \approx 0.39$ and 
the upper limit of 0.9 obtained from millisecond pulsars \citep{Kramer1998ApJ...501..270K}, 
this relation can be extended to the formula \citep{Titus2020MNRAS.494..500T}
\begin{equation}\label{eq:beam2}
f_{\rm TM98}(P \le 0.1~{\rm s}) \approx -0.255~\log (P/{\rm s}) + 0.135.
\end{equation}

\begin{table}
\label{MyTabD}
\small
\begin{center}
\centering \caption{Parameters of the Parkes multibeam pulsar survey.}
    \begin{tabular}{ll}
    \hline\hline
    Sampling interval ($t_{\rm samp}$) & 250~$\mu$s \\
    Receiver temperature ($T_{\rm r}$) & 21~K\\
    Number of channels ($N_{\rm ch}$) & 96 \\
    Antenna gain at the beam centre ($G_0$) & 0.7~K~Jy$^{-1}$ \\
    Number of polarizations ($N_{\rm p}$)& 2\\
    Receiver bandwidth ($\Delta_{\nu}$) & 288~MHz \\
    Full width at half maximum($w$) & 14~arcmin \\
    Integration time ($t_{\rm int}$) & 35~minutes \\
    \hline
    \end{tabular}
\end{center}
\end{table}

\clearpage
\bibliography{cite}{}
\bibliographystyle{aasjournal}

\end{document}